\documentclass[pre,a4paper,oneside,twocolumn]{revtex4}

\usepackage{graphicx}  
\usepackage{latexsym}   

\begin{document}

\title{Exact analytical calculation for the percolation crossover in deterministic partially self-avoiding walks in one-dimensional random media}

\author{\firstname{C\'esar} Augusto Sangaletti \surname{Ter\c{c}ariol}} 
\email{cesartercariol@gmail.com}
\affiliation{Faculdade de Filosofia, Ci\^encias e Letras de Ribeir\~ao Preto, \\
             Universidade de S\~ao Paulo \\ 
             Avenida Bandeirantes, 3900 \\ 
             14040-901, Ribeir\~ao Preto, SP, Brazil}

\affiliation{Centro Universit\'ario Bar\~ao de Mau\'a \\
             Rua Ramos de Azevedo, 423 \\ 
             14090-180, Ribeir\~ao Preto, SP, Brazil}

\author{\firstname{Rodrigo} Silva \surname{Gonz\'alez}}
\email{caminhos_rsg@yahoo.com.br}
\affiliation{Faculdade de Filosofia, Ci\^encias e Letras de Ribeir\~ao Preto, \\
             Universidade de S\~ao Paulo \\ 
             Avenida Bandeirantes, 3900 \\ 
             14040-901, Ribeir\~ao Preto, SP, Brazil}

\author{\firstname{Alexandre} Souto \surname{Martinez}}
\email{asmartinez@ffclrp.usp.br}

\affiliation{Faculdade de Filosofia, Ci\^encias e Letras de Ribeir\~ao Preto, \\
             Universidade de S\~ao Paulo \\ 
             Avenida Bandeirantes, 3900 \\ 
             14040-901, Ribeir\~ao Preto, SP, Brazil}
\date{\today}

\begin{abstract}
Consider $N$ points randomly distributed along a line segment of unitary length.
A walker explores this disordered medium moving according to a partially self-avoiding deterministic walk.
The walker, with memory $\mu$, leaves from the leftmost point and moves, at each discrete time step, to the nearest point which has not been visited in the preceding $\mu$ steps.
Using open boundary conditions, we have calculated analytically the probability $P_N(\mu) = (1 - 2^{-\mu})^{N - \mu - 1}$ that all $N$ points are visited, with  $N \gg \mu \gg 1$.
This approximated expression for $P_N(\mu)$ is  reasonable even for small $N$ and $\mu$ values, as validated by Monte Carlo simulations.
We show the existence of a critical memory $\mu_1 = \ln N/\ln 2$.
For $\mu < \mu_1 - e/(2\ln2)$, the walker gets trapped in cycles and does not fully explore the system. 
For $\mu > \mu_1 + e/(2\ln2)$ the walker explores the whole system.  
Since the intermediate region increases as $\ln N$ and its width is constant, a sharp transition is obtained for one-dimensional large systems.
This means that the walker needs not to have full memory of its trajectory to explore the whole system.
Instead, it suffices to have memory of order $\log_{2} N$.   
\end{abstract}



\maketitle

\section{Introduction}

While random walks in regular or disordered media have been thouroughly explored~\cite{fisher:1984}, deterministic walks in regular~\cite{grassberger:92} and disordered media~\cite{bunimovich:2004,boyer_2004,boyer_2005,boyer_2006} have been much less studied. 
Here we are concerned with the properties of deterministic walks in random media.

Given $N$ points distributed in a $d$-dimensional space, a possible question to ask is how efficiently these points can be visited by a walker who follows a simple movimentation rule. 
The search for the shortest closed path passing once in each point is the well known \emph{travelling salesman problem} (TSP), which has been extensively studied. 
In particular, if the points coordinates are distributed following a uniform deviate, results concerning the statistics of the shortest paths have been obtained analytically~\cite{percus:1996, percus:1997, percus:1999}. 
To tackle the TSP, one imperatively needs to know the coordinates of all the points in advance.
Global system information must be at the walker's disposal.

Nevertheless, other situations may be envisaged.
For instance, suppose that only local information about the neighborhood ranking of the current point is at the walker's disposal.
In this case, one can think of several deterministic and stochastic strategies to maximize the number of visited points while trying to minimize the travelled distance. 

Our aim is to study the way a walker explores the medium following the deterministic rule of going to the nearest point, which has not been visited in the previous $\mu$ discrete time steps. 
We call this partially self-avoiding walk of the \emph{deterministic tourist walk}.
Each trajectory, produced by this deterministic rule, has an initial transient of length $t$ and ends in a cycle of period $p$. 
Both transient time and cycle period can be combined in the joint distribution $S_{\mu, d}^{(N)}(t,p)$. 
The deterministic tourist walk with memory $\mu = 0$ is trivial.
Every starting point is its own nearest neighbor, so the trajectory contains only one point.
The transient and period joint distribution is simply $S^{(N)}_{0, d}(t, p) = \delta_{t,0} \delta_{p,1}$, where $\delta_{i,j}$ is the Kronecker delta. 
With memory $\mu = 1$, the walker must leave the current point at each time step.
The transient and period joint distribution has been obtained analytically for $N \gg 1$~\cite{tercariol_2005}.
This memoryless rule ($\mu = 1$) does not lead to exploration of the random medium since after a very short transient, the tourist gets trapped in pairs of points that are mutually nearest neighbors.
Interesting phenomena occur when greater memory values are considered. 
In this case, the cycle distribution is no longer peaked at $p_{min} = \mu + 1$, but presents a whole spectrum of cycles with period $p \ge p_{min}$, with possible power-law decay~\cite{lima_prl2001,stanley_2001,kinouchi:1:2002}. 
These cycles have been used as a clusterization method~\cite{campiteli_2006} and in image texture analysis~\cite{backes_2006,bruno_2006}. 

It is interesting to point out that, for 1D systems, determinism imposes serious restrictions.
For any $\mu$ value, cycles of period $2\mu+1 \le p \le 2\mu+3$ are forbidden.
Additionally, for $\mu = 2$ all odd periods but $p_{min} = 3$ are forbidden.
Also, the heavy tail of the period marginal distribution $S_{\mu, 1}^{(N)}(p) = \sum_t S_{\mu, 1}^{(N)}(t,p)$ may lead to often-visited-large-period cycles~\cite{lima_prl2001}.
This allows system exploration even for small memory values ($\mu \ll N$).

The article presentation is divided as follows.
In Sec.~\ref{semi-infinite_medium}, we consider a walker moving according to the deterministic tourist rule in  
semi-infinite disordered media. 
Firstly, we calculate exactly the distribution of visited points, which allowed us to jusfify a very good approximation using a simple mean field argument. 
Secondly, we propose an alternative exact derivation for this distribution using the exploration and return probabilities, which allows application in the tourist walk in finite disordered media. 
This is done in Sec.~\ref{finite_medium}, where we obtain the percolation probability and show the existence of a crossover in the walker's exploratory behavior at a critical memory $\mu_1 = \ln N / \ln 2$ in a narrow memory range of width $\varepsilon = e/\ln 2$. 
This crossover splits the walker's behavior in essentially two regimes.
For $\mu < \mu_1 - \varepsilon/2$, the walker gets trapped in cycles and for $\mu > \mu_1 + \varepsilon/2$, the walker visits all the points. 
The calculated quantities have been validated by Monte Carlo simulations. 
The fact that to explore the whole disordered medium the walker need to have only a small memory (of order $\log_2 N$) and other final remarks are presented in Sec.~\ref{conclusion}.  

\section{Semi-infinite disordered media}
\label{semi-infinite_medium}

\begin{figure}[htb]
\begin{center}
\includegraphics[angle=0,width=\columnwidth]{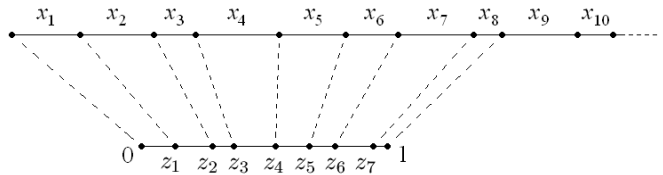}
\caption{Scheme showing the equivalence between a finite and a semi-infinite disordered media. 
         Along the upper line segment the points are generated using the random distances $x_k$ with exponential pdf.
         In the lower line segment, the number of points ($N$) is fixed and normalized to its total length, where $z_k$ are the normalized coordinates.}
\label{Fig:Equivalencia}
\end{center}
\end{figure}

A random static semi-infinite medium is constructed by uncountable points that are randomly and uniformly distributed along a semi-infinite line segment with a mean density $r$.
The upper line segment of Fig.~\ref{Fig:Equivalencia} represents this medium, where the distances $x_k$ between consecutive points are independent and identically distributed (iid) variables with exponential probability density function (pdf): $g(x) = r e^{-r x}$, for $x \ge 0$ and $g(x) = 0$, otherwise.
In the following we analytically obtain the statistics related to deterministic tourist walk performed on semi-infinite random media.

\subsection{Distribution of the number of visited points}

Here we obtain analytically the probability $S_{\mu, si}^{(\infty)}(n)$ for a walker, with memory $\mu$ and moving according to the deterministic tourist rule, to visit $n$ points of semi-infinite media.
The exact result is obtained and this allowed us to justify a simple mean field approach. 

\subsubsection{Exact Result}

Consider a walker who leaves from the leftmost point $s_1$, placed at the origin of the upper line segment of Fig.~\ref{Fig:Equivalencia}.
The conditions for the walker to visit $n \ge \mu+1$ distinct points are:
\begin{enumerate}
\item the distances $x_1$, $x_2$, \ldots, $x_\mu$ may assume any value in the interval $[0; \infty)$, since the memory $\mu$ prohibits the walker to move backwards in the first $\mu$ steps, so that the first $\mu+1$ points are indeed visited,  
\item  each of the following distances $x_{\mu+1}$, $x_{\mu+2}$, \ldots, $x_{n-1}$ must be smaller than the sum of the $\mu$ preceding step distances, until the tourist reaches the point $s_n$, and 
\item the distance $x_n$ must be greater than the sum of the $\mu$ preceding ones, to enforce the walker to move back to the point $s_{n-\mu}$, instead of exploring a new point $s_{n+1}$.
\end{enumerate}
Once the walker has returned to the point $s_{n-\mu}$, he/she may revisit the starting point $s_1$, get trapped in a attrator or even revisit the point $s_n$, but he/she will not be able to transpose the distance barrier $x_n$ between the points $s_n$ and $s_{n+1}$.
Actually, no new points will be visited any longer.
Combining these conditions, the probability for the walker to visit $n$ distinct points is
\begin{eqnarray}
\label{Eq:DistrX}
S_{\mu, si}^{(\infty)}(n)
& = & \prod_{j=1}^\mu \int_0^\infty \mbox{d}x_j \; r e^{-r x_j} \nonumber \\
&   & \prod_{j=\mu+1}^{n-1} \int_0^{\sum_{k=j-\mu}^{j-1} x_k} \mbox{d}x_j \; r e^{-r x_j} \nonumber \\
&   & \int_{\sum_{k=n-\mu}^{n-1} x_k}^\infty \mbox{d}x_n \; r e^{-r x_n} \; .
\end{eqnarray}
The difficulty to obtain $S_{\mu, si}^{(\infty)}(n)$ is that the $n$ intregrals are chained and the integration procedure must start from the rightmost factor. 
Applying the substitutions $y_j = e^{-r x_j}$, with  $1 \le j \le n$, one has:
\begin{eqnarray}
\label{Eq:DistrYAbrev}
S_{\mu, si}^{(\infty)}(n) & = & \prod_{j=1}^n {\cal I}_{j}  \; ,
\end{eqnarray}
where the form of each functional ${\cal I}_j$ depends on $j$: 
\begin{eqnarray}
\label{Eq:Ij}
{\cal I}_{j} = 
\left\{
\begin{array}{ll}
\int_0^1 \mbox{d} y_j \; , & \mbox{for} \; 1 \le j \le \mu \\ \\
\int_{\tilde{y}_j}^1 \mbox{d} y_j \; , & \mbox{for} \; \mu + 1 \le j \le n - 1 \\ \\
\int_0^{\tilde{y}_j} \mbox{d} y_j \; , & \mbox{for} \; j = n \\
\end{array}
\right.
\end{eqnarray}
and each integration limit $\tilde{y}_j = \prod_{k=j-\mu}^{j-1} y_k$ links ${\cal I}_j$ to the preceding $\mu$ integrals. 
This means that Eq.~\ref{Eq:DistrYAbrev} must be evaluated from ${\cal I}_n$ to ${\cal I}_1$.
Notice that $r$ has been eliminated, indicating that the number of visited points does not depend on the medium density.

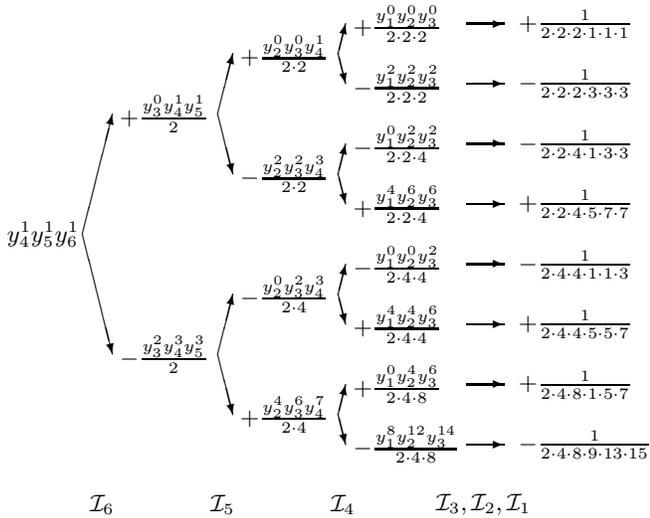
\begin{figure}[htb]
\begin{center}
\setlength{\unitlength}{1mm}
\begin{picture}(90, 68)

\put(0, 36){\makebox(0, 0)[l]{$y_4^1 y_5^1 y_6^1$}}

\put(10, 36){\vector(1, 4){4}}
\put(10, 36){\vector(1, -4){4}}
\put(15, 52){\makebox(0, 0)[l]{$+\frac{y_3^0 y_4^1 y_5^1}{2}$}}
\put(15, 20){\makebox(0, 0)[l]{$-\frac{y_3^2 y_4^3 y_5^3}{2}$}}

\put(11, 0){\makebox(0, 0)[l]{${\cal I}_6$}}

\multiput(28, 20)(0, 32){2}{\vector(1, 4){2}}
\multiput(28, 20)(0, 32){2}{\vector(1, -4){2}}
\put(31, 60){\makebox(0, 0)[l]{$+\frac{y_2^0 y_3^0 y_4^1}{2 \cdot 2}$}}
\put(31, 44){\makebox(0, 0)[l]{$-\frac{y_2^2 y_3^2 y_4^3}{2 \cdot 2}$}}
\put(31, 28){\makebox(0, 0)[l]{$-\frac{y_2^0 y_3^2 y_4^3}{2 \cdot 4}$}}
\put(31, 12){\makebox(0, 0)[l]{$+\frac{y_2^4 y_3^6 y_4^7}{2 \cdot 4}$}}

\put(27, 0){\makebox(0, 0)[l]{${\cal I}_5$}}

\multiput(44, 12)(0, 16){4}{\vector(1, 4){1}}
\multiput(44, 12)(0, 16){4}{\vector(1, -4){1}}
\put(46, 64){\makebox(0, 0)[l]{$+\frac{y_1^0 y_2^0    y_3^0   }{2 \cdot 2 \cdot 2}$}}
\put(46, 56){\makebox(0, 0)[l]{$-\frac{y_1^2 y_2^2    y_3^2   }{2 \cdot 2 \cdot 2}$}}
\put(46, 48){\makebox(0, 0)[l]{$-\frac{y_1^0 y_2^2    y_3^2   }{2 \cdot 2 \cdot 4}$}}
\put(46, 40){\makebox(0, 0)[l]{$+\frac{y_1^4 y_2^6    y_3^6   }{2 \cdot 2 \cdot 4}$}}
\put(46, 32){\makebox(0, 0)[l]{$-\frac{y_1^0 y_2^0    y_3^2   }{2 \cdot 4 \cdot 4}$}}
\put(46, 24){\makebox(0, 0)[l]{$+\frac{y_1^4 y_2^4    y_3^6   }{2 \cdot 4 \cdot 4}$}}
\put(46, 16){\makebox(0, 0)[l]{$+\frac{y_1^0 y_2^4    y_3^6   }{2 \cdot 4 \cdot 8}$}}
\put(46,  8){\makebox(0, 0)[l]{$-\frac{y_1^8 y_2^{12} y_3^{14}}{2 \cdot 4 \cdot 8}$}}

\put(43, 0){\makebox(0, 0)[l]{${\cal I}_4$}}

\multiput(61, 8)(0, 8){8}{\vector(1, 0){5}}
\put(68, 64){\makebox(0, 0)[l]{$+\frac{1}{2 \cdot 2 \cdot 2 \cdot 1 \cdot  1 \cdot  1}$}}
\put(68, 56){\makebox(0, 0)[l]{$-\frac{1}{2 \cdot 2 \cdot 2 \cdot 3 \cdot  3 \cdot  3}$}}
\put(68, 48){\makebox(0, 0)[l]{$-\frac{1}{2 \cdot 2 \cdot 4 \cdot 1 \cdot  3 \cdot  3}$}}
\put(68, 40){\makebox(0, 0)[l]{$+\frac{1}{2 \cdot 2 \cdot 4 \cdot 5 \cdot  7 \cdot  7}$}}
\put(68, 32){\makebox(0, 0)[l]{$-\frac{1}{2 \cdot 4 \cdot 4 \cdot 1 \cdot  1 \cdot  3}$}}
\put(68, 24){\makebox(0, 0)[l]{$+\frac{1}{2 \cdot 4 \cdot 4 \cdot 5 \cdot  5 \cdot  7}$}}
\put(68, 16){\makebox(0, 0)[l]{$+\frac{1}{2 \cdot 4 \cdot 8 \cdot 1 \cdot  5 \cdot  7}$}}
\put(68,  8){\makebox(0, 0)[l]{$-\frac{1}{2 \cdot 4 \cdot 8 \cdot 9 \cdot 13 \cdot 15}$}}

\put(57, 0){\makebox(0,0)[l]{${\cal I}_3, {\cal I}_2, {\cal I}_1$}}

\end{picture}
\caption{Calculation scheme for the chained integrals of Eq.~\ref{Eq:DistrYAbrev}. 
         Here we have considered the example of $\mu = 3$ and $n = 7$. 
         We focus on the dynamics of the powers of $y$'s along the bifurcation path, which leads to the recursive relation (Eq.~\ref{Eq:F}).} 
\label{Fig:CalcIntEncad}
\end{center}
\end{figure}

We call attention that the novelty of this calculation concerns dealing with the powers of $y_j$. 
Fig.~\ref{Fig:CalcIntEncad} illustrates the calculation of Eq.~\ref{Eq:DistrYAbrev} for the particular case $\mu=3$ and $n=7$. 
In this scheme, the relevant quantities are the $y_j$ powers in the integrand, since all $y$'s disappear after all integration levels are performed. 
The integration process consists basicly of three steps, where each one of them represents a case of Eq.~\ref{Eq:Ij}.
\begin{enumerate}
\item the first integral ${\cal I}_7$ (third case of Eq.~\ref{Eq:Ij}) is trivially evaluated to its upper limit $\tilde{y}_7$, yielding the root node $y_4^1 y_5^1 y_6^1$, with all the integrand variables raised to the first power.
These powers are denoted as $a_1$, $a_2$, \ldots, $a_{\mu}$, where, in particular, $a_{\mu}$ is the power of integrand of the current level.
\item each bifurcation level represents an integral from ${\cal I}_6$ to ${\cal I}_4$ (second case of Eq.~\ref{Eq:Ij}).
\begin{enumerate}
\item a unit is added to the power $a_{\mu}$ and it becomes a new factor $a_{\mu} + 1$ at the denominator of the following level [this is just $\int y^a \; \mbox{d}y = y^{a+1}/(a+1)$].
\item for each bifurcation, in the upper fractions, the remaindering variables keep their powers and the new $y$ raised to 0.
\item in the lower fraction, we sum $a_{\mu} + 1$ to each power of $y$'s and fraction sign is switched.
\end{enumerate}
\item the last level represents the integrals from ${\cal I}_3$ to ${\cal I}_1$ (first case of Eq.~\ref{Eq:Ij}), where a unit is added to all powers $a_1$, $a_2$, \ldots, $a_\mu$ and they become new factors at the denominator.
\end{enumerate}

Generalizing the reasoning of the scheme of Fig.~\ref{Fig:CalcIntEncad} for arbitrary $\mu$ and $n$, Eq.~\ref{Eq:DistrYAbrev} may be writen as the following recursive formula
\begin{eqnarray}
\label{Eq:DistRec}
S_{\mu, si}^{(\infty)}(n) & = & f_\mu[n, \vec{1}] \; , \hspace{6mm} n = \mu+1, \mu+2, \ldots, \infty \; ,
\end{eqnarray}
with
\begin{widetext}
\begin{eqnarray}
\label{Eq:F}
f_\mu[j, \vec{a}] = \left\{ \begin{array}{ll}
\frac{f_\mu[j-1, \; \mbox{{\scriptsize shift}}(\vec{a})] - f_\mu[j-1, \; \mbox{{\scriptsize shift}}(\vec{a})+(a_\mu+1) \cdot \vec{1}]}{a_\mu+1} \; , & \mbox{if } j>\mu+1 \\
1/\prod_{k=1}^\mu (a_k+1) \; , & \mbox{if } j=\mu+1 \\
\end{array} \right.
\end{eqnarray}
\end{widetext}
where $\vec{1} = (1, 1, 1, \ldots, 1)$ and $\vec{a} = (a_1, a_2, a_3, \ldots, a_\mu)$ are $\mu$-dimensional vectors, $\mbox{shift}(\vec{a}) = (0, a_1, a_2, \ldots, a_{\mu-1})$ is the aciclic coordinates shifting and $j$ is the integration level, also used as stop condition.
Observe that the initial condition $\vec{1}$ of Eq.~\ref{Eq:DistRec} and the upper and lower cases of Eq.~\ref{Eq:F} represent the third, second and first cases of Eq.~\ref{Eq:Ij}, respectively.

Observe that $p_{min} = \mu+1$ is the minimum allowed cycle period in the deterministic tourist walk~\cite{lima_prl2001} and, once the memory $\mu$ assures the walker visits at least $\mu+1$ points, the number of extra visited points $n_e = n- p_{min}$ is the relevant quantity since all the  distributions $S_{\mu, si}^{(\infty)}$ start at the same point $n_e=0$, regardless the $\mu$ value.

\begin{figure}[htb]
\begin{center}
\includegraphics[angle=0,width=\columnwidth]{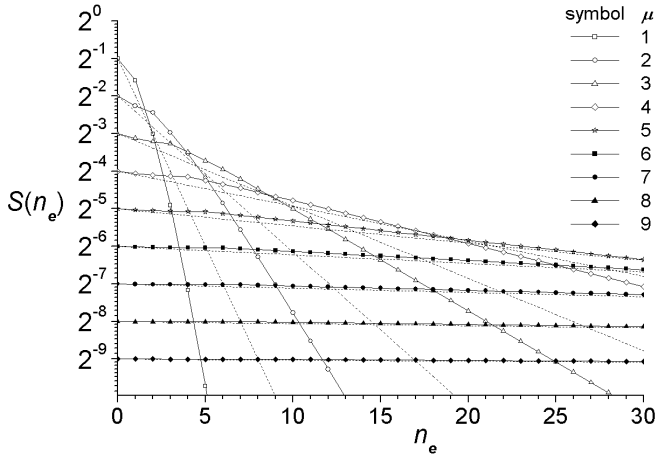}
\caption{Distribution of $n_e$ for $\mu$ varying from 1 to 9.
Continuous lines refer to exact form of Eq.~\ref{Eq:DistRec} and dotted lines refer to approximated form of Eq.~\ref{Eq:DistrNeAprox}.}
\label{Fig:DistrNe}
\end{center}
\end{figure}

Although the recursive relation of Eq.~\ref{Eq:F} is exact, it is not efficient for algebrical treatment.
Even for numerical calculation it presents several disadvantages.
It is difficult to implement due to its recursive structure and the processing time grows exponentially.
Such exponential time-dependence limited the plots of Fig.~\ref{Fig:DistrNe} and Fig.~\ref{Fig:ProbRecuo} to $n_e \le 30$.
The continous lines of Fig.~\ref{Fig:DistrNe} represent Eq.~\ref{Eq:DistRec} for different values of $\mu$. 
As one can see from this figure, a remarkable property is 
\begin{equation}
S_{\mu, si}^{(\infty)}(n_e=0) = \frac{1}{2^\mu} \; ,
\end{equation}
for all $\mu$. 
This is exactly the probability to have a null transient and a cycle with minimum period $p_{min}$ in the one-dimensional tourist walk. 

\subsubsection{Mean field approximation}

The recursivity of Eq.~\ref{Eq:F} has been inherited from the chained integrals of Eq.~\ref{Eq:DistrYAbrev}.
However, for $\mu \gg 1$ a mean field approximation may be used to untie those integrals.
It consists of replacing the products $\tilde{y}_j$ by their mean values. 

To fully appreaciate this mean field argumentation consider first the distribution of a product of uniform deviates. 
Let $y_1$, $y_2$, \ldots, $y_\mu$ be $\mu$ independent random variables uniformly distributed on the interval $(0, 1]$.
To obtain the pdf $p(\tilde{y})$ of the product $\tilde{y} = \prod_{k=1}^\mu y_k$, let us apply the transformation $\tilde{w} = -\ln \tilde{y} = \sum_{k=1}^\mu w_k$, where $w_k = -\ln y_k$ with $1 \le k \le \mu$ are iid variables with exponential pdf of unitary mean.
Thus, the sum $\tilde{w}$ follows a gamma pdf $p(\tilde{w}) = \tilde{w}^{\mu-1} e^{-\tilde{w}}/\Gamma(\mu)$.
Since $|p(\tilde{y}) \mbox{d}\tilde{y}| = |p(\tilde{w}) \mbox{d}\tilde{w}|$ one obtains the distribution of $\tilde{y}$: $p(\tilde{y}) = (-\ln \tilde{y})^{\mu-1}/\Gamma(\mu)$, whose the $m$th moment is $\langle \tilde{y}^m \rangle = (m+1)^{-\mu}$.

The above tools can be used due to the fact that all the variables $y_j = e^{-r x_j}$ (applyed to Eq.~\ref{Eq:DistrX}) are iid according to a uniform deviate in the interval $(0, 1]$.
The first condition ($0 \le j \le \mu$) of Eq.~\ref{Eq:Ij} states that the variables $y_1$, $y_2$, \ldots, $y_\mu$ may freely vary from 0 to 1.
Once for $\mu \gg 1$ the product $\tilde{y}_{\mu+1}=\prod_{k=1}^\mu y_k$ has a small variance, it can be approximated by its mean value $\langle \tilde{y}_{\mu+1} \rangle = 2^{-\mu}$.

Concerning the next product $\tilde{y}_{\mu+2}=\prod_{k=2}^{\mu+1} y_k$, the variables $y_2$, $y_3$, \ldots, $y_{\mu+1}$ are not all iid, because $y_{\mu+1}$ has just been constrained to the interval $[2^{-\mu}, 1]$.
However, for $\mu \gg 1$, the interval $[2^{-\mu}, 1]$ becomes close to $[0, 1]$, allowing $\tilde{y}_{\mu+2}$ to be also approximated by the mean value $2^{-\mu}$.
This reasoning can be indutively applied for the remaining integration limits $\tilde{y}_j$.
Thus, Eq.~\ref{Eq:Ij} is approximated to
\begin{eqnarray}
\label{Eq:IjAprox}
{\cal I}_{j} \approx 
\left\{
\begin{array}{ll}
\int_0^1 \mbox{d} y_j \; , & \mbox{for} \; 0 \le j \le \mu \\ \\
\int_{2^{-\mu}}^1 \mbox{d} y_j \; , & \mbox{for} \; \mu + 1 \le j \le n - 1 \\ \\
\int_0^{2^{-\mu}} \mbox{d} y_j \; , & \mbox{for} \; j = n \\ \\
\end{array}
\right. \; .
\end{eqnarray}
Observe that these integrals are no longer chained and that $S_{\mu, si}^{(\infty)}(n)$ is still given by Eq.~\ref{Eq:DistrYAbrev}, which leads to
\begin{eqnarray}
\label{Eq:DistrNeAprox}
S_{\mu, si}^{(\infty)}(n) \approx 2^{-\mu} (1-2^{-\mu})^{n-\mu-1} \; ,
\end{eqnarray}
with $n = \mu+1, \mu+2, \ldots, \infty$ and yields $\mbox{E}(n) = 2^\mu + \mu$, which may be interpreted as the {\em characteristic range} of the walk, and $\mbox{Var}(n) = 2^{2\mu}-2^\mu$.
Dotted lines in Fig.~\ref{Fig:DistrNe} represent this approximation for $1 \le \mu \le 9$.

\subsection{Exploration and return probabilities}

The purpose of the calculation of the exploration and return probabilities is twofold. 
It is an alternative argumentation to obtain Eq.~\ref{Eq:DistrNeAprox} and these probabilities lead to simple arguments to obtain the percolation probability for a finite disordered. 

\subsubsection{Upper tail cumulative probability: an exact calculation}

A similar argumentation used to obtain Eq.~\ref{Eq:DistRec} may be used to obtain the upper tail cummulative distribution $\overline{F}_{\mu, si}^{(\infty)}(n) = \sum_{k=n}^\infty S_{\mu, si}^{(\infty)}(k)$.
This distribution gives the probability for the walker to visit at least $n$ points.
The only modification is that, once the walker has reached the point $s_n$, he/she can either move backwards or forwards.
Therefore, the rightmost integral of Eq.~\ref{Eq:DistrX} is no longer necessary, so
\begin{eqnarray}
\label{Eq:FProdI}
\overline{F}_{\mu, si}^{(\infty)}(n) & = & \prod_{j=1}^{n-1} {\cal I}_{j}  \; ,
\end{eqnarray}
where each functional ${\cal I}_j$ is given by Eq.~\ref{Eq:Ij}.
The root node of Fig.~\ref{Fig:CalcIntEncad} is now set to 1 (or, equivalently, $y_4^0 y_5^0 y_6^0$), which leads to
\begin{eqnarray}
\label{Eq:DistAcum}
\overline{F}_{\mu, si}^{(\infty)}(n) = f_\mu[n, \vec{0}] \; , \hspace{6mm} n = \mu+1, \mu+2, \ldots, \infty \; ,
\end{eqnarray}
where $\vec{0} = (0, 0, \ldots, 0)$ is the $\mu$-dimensional null vector and $f_\mu$ is given by Eq.~\ref{Eq:F}. 
Observe that $\overline{F}_{\mu, si}^{(\infty)}(n)$ uses the same recursive structure of Eq.~\ref{Eq:DistRec}, but with a different initial condition ($\vec{0}$ instead of $\vec{1}$). 
If the approximation of Eq.~\ref{Eq:IjAprox} is used as approximation to evaluate Eq.~\ref{Eq:FProdI}, one readily has
\begin{eqnarray}
\label{Eq:DistrAcumAprox}
\overline{F}_{\mu, si}^{(\infty)}(n) \approx (1-2^{-\mu})^{n-\mu-1} \; .
\end{eqnarray}

The memory $\mu$ assures the walker, leaving from the point $s_1$, to move forward in the first $\mu$ steps.
In constrast, the following steps are uncertain, since the walker may either move forwards and visit a new point or return and stop the medium exploration.
In analogy to the geometric distribution, it is useful to define the exploration probability $q_\mu(j)$ (taken as failure) as the probability for the walker to visit a new point at the $j$th uncertain step.

Therefore, the return probability $p_\mu(j)$ (taken as success) for the $j$th uncertain step is equal the probability for the walker to visit exactly $n=\mu+j$ points conditionated to the fact that he/she has already visited $n=\mu+j-1$ points.
This probability is given by
\begin{eqnarray}
\label{Eq:ReturnProb}
p_\mu(j) = \frac{S_{\mu, si}^{(\infty)}(n=\mu+j)}{\overline{F}_{\mu, si}^{(\infty)}(n=\mu+j)} = \frac{f_\mu[\mu+j, \vec{1}]}{f_\mu[\mu+j, \vec{0}]} \; ,
\end{eqnarray}
where $f_\mu$ is given by Eq.~\ref{Eq:DistRec}.

\begin{figure}[htb]
\begin{center}
\includegraphics[angle=0,width=\columnwidth]{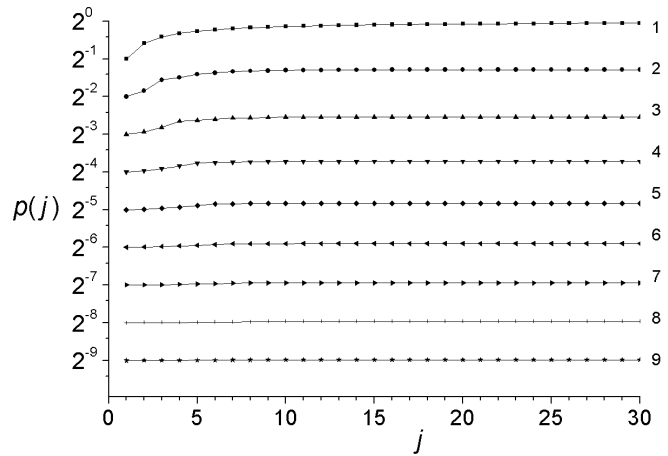}
\caption{Return probability given by Eq.~\ref{Eq:ReturnProb}, with $\mu$ varying from 1 to 9.}
\label{Fig:ProbRecuo}
\end{center}
\end{figure}

Fig.~\ref{Fig:ProbRecuo} shows the behavior of $p_\mu(j)$ for the first 30 uncertain steps, with $\mu$ varying from 1 to 9.
One can observe that for $\mu \gg 1$ the return probability $p_\mu(j)$ along the walk is almost constant and equal its initial value $p_\mu(1)=2^{-\mu}$.
In this way, one can verify empirically that for $\mu \gg 1$ the return probabilities can be taken as $p_\mu=2^{-\mu}$ for all steps, as well as $q_\mu=1-2^{-\mu}$ can be taken for all exploration probabilities.

This empirical approximation for the return probability can justified analytically using Eqs.~\ref{Eq:DistrNeAprox}~and~\ref{Eq:DistrAcumAprox} in its definition:
\begin{eqnarray}
\label{Eq:ReturnProbAprox}
p_\mu(j) = \frac{S_{\mu, si}^{(\infty)}(n=\mu+j)}{\overline{F}_{\mu, si}^{(\infty)}(n=\mu+j)} \approx \frac{1}{2^\mu} \; ,
\end{eqnarray}
 
For $\mu=1$, $n_e$ is numerically equal to the transient time $t$ (what does not mean that they are the same part of the trajectory, the transient is the beginning of it and $n_e$ counts the final points), and in this case Eqs.~\ref{Eq:DistRec}, ~\ref{Eq:DistAcum} and \ref{Eq:ReturnProb} assume the simple exact closed forms $S_{1, si}^{(\infty)}(n_e) = (n_e+1)/(n_e+2)!$, $\overline{F}_{1, si}^{(\infty)}(n_e) = 1/(n_e+1)!$ and $p_1(j)=j/(j+1)$, which have been previously found in Ref.~\cite{tercariol_2005}.

\subsubsection{An alternative derivation}

These approximated expressions for exploration and return probabilities can also be obtained by analytical means through a more direct derivation.
Consider again the tourist dynamics with a walker who leaves from the point $s_1$, placed at the origin of the semi-infinte medium.

The first $\mu+1$ points are indeed visited, because the memory $\mu$ prohibits the walker to return.
Thus, the distances $x_1$, $x_2$, \ldots, $x_\mu$ may assume any value in the interval $[0, \infty)$.

The exploration probability $q_\mu(1)$ for the first uncertain step can be obtained imposing that the distance $x_{\mu+1}$ must be smaller than the sum $\tilde{x}_1=\sum_{k=1}^\mu x_k$.
Since the variables $x_1$, $x_2$, \ldots, $x_\mu$ are iid with exponential pdf, $\tilde{x}_1$ has a gamma pdf.
Hence $q_\mu(1) = [\int_0^\infty \mbox{d}\tilde{x}_1 r^\mu \tilde{x}_1^{\mu-1} e^{-r \tilde{x}_1} /\Gamma(\mu)] \int_0^{y_1} \mbox{d}x_{\mu+1} re^{-rx_{\mu+1}} = 1-2^{-\mu}$. 

The exploration probability $q_\mu(2)$ for the second uncertain step is not exactly equal to $q_\mu(1)$.
Once the distance $x_{\mu+1}$ must vary in the interval $[0, \tilde{x}_1]$, the variables $x_2$, $x_3$, \ldots, $x_{\mu+1}$ are not all independent, and consequently $\tilde{x}_2=\sum_{k=2}^{\mu+1} x_k$ has not exactly a gamma pdf.
However, for $\mu \gg 1$, $x_{\mu+1}$ rarely exceeds $\tilde{x}_1$ [this probability is just $P(x_{\mu+1}>\tilde{x}_1) = 1-q_\mu(1)=2^{-\mu}$, meaning that a weak correlation is present for $\mu \gg 1$]. 
Therefore, one can make an approximation assuming that $\tilde{x}_2$ still follows a gamma pdf and considering $q_\mu(2) \approx q_\mu(1)$.
The same argumentation can be used for the succeeding steps.

When the point $s_n$ is reached, the walker must turn back, stopping the medium exploration.
Once $q_\mu(1)$ is taken for all $q_\mu$, the return probability is $p_\mu = 1-q_\mu = 2^{-\mu}$ and one has: $S_{\mu, si}^{(\infty)}(n_e) = 2^{-\mu} (1-2^{-\mu})^{n_e}$, which is the result of Eq.~\ref{Eq:DistrNeAprox}.

\section{Percolation probability for finite disordered media}
\label{finite_medium}

The finite disordered medium is constructed by $N$ points whose coordinates $z_k$ are randomly generated in the interval $[0, 1]$ following a uniform deviate as depicted in Fig.~\ref{Fig:Equivalencia}.

Numerical simulation results pointed out that the exploration and return probabilities obtained for the semi-infinite medium may also be applied to this finite medium.
This is not trivial, since all results for the semi-infinite medium have been obtained assuming that the distances between consecutive points are iid variables with exponential distribution.
Obviously the distances between consecutive points in the finite medium are not iid variables, nor have exponential distribution.

Nevertheless, the equivalence between these two media can be argued as follows.
The abscissas of the ranked points in the finite medium follow a beta pdf~\cite{mathworld:gamma}.
If one restricts the semi-infinite medium length to the first $N+1$ distances and normalizes it to fit in the interval $[0, 1]$, then the abscissa of its $k$th ranked point is $z_k=U_k/(U_k+V_k)$, where $U_k=x_1+x_2+\cdots+x_k$ and $V_k=x_{k+1}+x_{k+2}+\cdots+x_{N+1}$.
Fig.~\ref{Fig:Equivalencia} shows an example for $N=7$ normalization.

This normalization does not affect the tourist walk, because in this walk only the neighborhood ranking is relevant, not the distances themselves~\cite{lima_prl2001,kinouchi:1:2002}.
Since $U_k$ and $V_k$ have gamma pdf, $z_k$ has beta pdf~\cite{mathworld:gamma}, as in the genuine finite medium.

The probability $P_N(\mu)$ for the exploration of the whole $N$-point medium can be derived noticing that the walker must move forward $N-(\mu+1)$ uncertain steps and, when the last point $s_N$ is reached, there is no need to impose a return step.
Therefore the percolation probability is 
\begin{eqnarray}
\label{Eq:Percolacao}
P_N(\mu) = q_\mu^{N-(\mu+1)} = \left(1-2^{-\mu}\right)^{N-\mu-1} \; .
\end{eqnarray}
It is interesting to point out that the percolation probability relates directly to the upper tail cummulative function as seen by Eq.~\ref{Eq:DistrAcumAprox}.
The diference between them is only on in the interpretation of the number of visited points $N$, but this can be justified because of the normalization to the finite medium discussed above.

\begin{figure}[htb]
\begin{center}
\includegraphics[angle=0,width=\columnwidth]{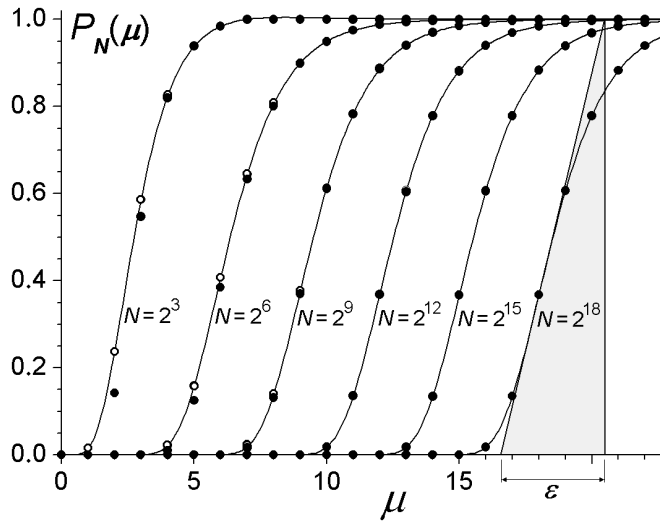}
\caption{Percolation probability for some fixed $N$ values.
Empty circles are given by Eq.~\ref{Eq:Percolacao} and full ones represent numerical simulations ($M=100\,000$ maps for each $N$ and $\mu$ values), error bars are smaller than symbol size.
Continuous lines are plotted only to guide eyes.
Analytical results are satisfactory, when compared to numerical simulation, even for small $N$ and $\mu$ values. 
The crossover points $\mu_1$ are given by Eq.~\ref{Eq:MemCritica}, which are weakly dependent on $N$ but all of them have the same constant dispersion $\varepsilon \sim 4$ (Eq.~\ref{Eq:RegTrans}).}
\label{Fig:Percolacao}
\end{center}
\end{figure}

Fig.~\ref{Fig:Percolacao} shows a comparison of the evaluation of Eq.~\ref{Eq:Percolacao} and the results of Monte Carlo simulations. 
From this figure one clearly sees that the probability of full exploration increases abruptly from almost 0 to almost 1. 

From the analogy with a first order phase transition, we define the crossover point as the maximum of the derivative of $P_N(\mu)$, with respect $\mu$. 
This implies that the second derivative vanishes at the maximum $\mbox{d}_{\mu}^2 P_N(\mu)|_{\mu_1^{(c)}} = 0$, leading to a transcental equation which cannot be solved it analytically to obtain $\mu_1^{(c)}$. 
An estimation value of $\mu_1^{(c)}$ can be calculated considering $N \gg \mu \gg 1$ and Eq.~\ref{Eq:Percolacao} may be approximated to $P_N(\mu) = (1-2^{-\mu})^N$, and at inflexion point, one has
\begin{eqnarray}
\label{Eq:MemCritica}
\mu_1 & = & \log_2 N \; .
\end{eqnarray}
A simple interpretation can be given to $\mu_1$.
It is just the number of necessary bits to represent the system size $N$. 
To evaluate the width of the crossover region, use the slope of $P_N(\mu)$ at $\mu_1$, which results to $\ln 2/e$, for all $N$ [see Fig.~\ref{Fig:Percolacao}].
The crossover region has a constant width
\begin{eqnarray}
\label{Eq:RegTrans}
\varepsilon = \frac{e}{\ln 2} \approx 3.92 \; .
\end{eqnarray}
In one hand, as $N$ increases, the critical memory slowly increases (as $\log_2 N$) but its deviation is independent of the system size, so that a sharp crossover is found in the thermodynamic limit ($N \gg 1$).
We stress that the approximations employed lead to satisfactory results even for small $N$ and $\mu$ values. 

On the other hand, if one use the reduced memory $\tilde{\mu} = (\mu - \mu_1)/\mu_1$, the crossover occurs in $\tilde{\mu}_1 = 0$, but now the crossover width depends on the size of the system as $1/\log_2 N$. 

\section{Conclusion}
\label{conclusion}

Our main result is that to explore the whole medium the walker does not need to have memory of order $N$,  a small memory (of order $\ln N$) allows this full exploration. 

All the exact results calculated here are in accordance to the limiting case $\mu = 1$ obtained in Ref.~\cite{tercariol_2005}.
Also, they can be applied to an infinite line segment, where random points are distributed in both sides of the  starting point.  

An interesting result we have obtained in the one-dimensional deterministic tourist walk is that the probability to have a null transient and a minimum cycle is $2^{-\mu}$, where $\mu < \mu_1$ is the memory of the walker.  

The distance constraints can be generalized to a $d$-dimensional Euclidean space and possibly this calculation scheme can be employed in such interesting situation.  

Finally, the tourist rule can be relaxed to a stochastic walk.
In this case, the walker goes to nearer cities with greater probabilities, given by an one-parameter (inverse of the temperature) exponential distribution. 
This situation has been studied for the non-memory cases ($\mu = 0$~\cite{martinez:1:2004} and $\mu = 1$~\cite{risaugusman:1:2003}) and we have detected the existence of a critical temperature separating the localized from the extended regimes. 
It would be interesting  to combine both in the tourist walks, stochastic movimentation (driven by a temperature parameter) and memory ($\mu$).

\section*{Acknowledgements}

The authors thank N. A. Alves and F. M. Ramos for fruitful discussions.
ASM acknowledges the Brazilian agencies CNPq (305527/2004-5) and FAPESP (2005/02408-0) for support.


\end{document}